\documentclass[fleqn,usenatbib]{mnras}

\usepackage{newtxtext,newtxmath}

\usepackage[T1]{fontenc}
\usepackage{soul}

\DeclareRobustCommand{\VAN}[3]{#2}
\let\VANthebibliography\thebibliography
\def\thebibliography{\DeclareRobustCommand{\VAN}[3]{##3}\VANthebibliography}

\usepackage{graphicx}	
\usepackage{amsmath}	
\usepackage{amssymb}	

\newcommand{\angstrom}{\mbox{\normalfont\AA}}
\newlength\mylength
\title[Study of Recent outburst in RX J0209.6-7427 with \textit{AstroSat}]{Study of Recent outburst in the Be/X-ray binary RX J0209.6-7427 with \textit{AstroSat}: A new ultraluminous X-ray pulsar in the Magellanic Bridge?}

\author[A. D. Chandra et al.]{
Amar Deo Chandra,$^{1}$\thanks{E-mail: amar.deo.chandra@gmail.com}
Jayashree Roy,$^{2}$
P. C. Agrawal$^{3}$
and Manojendu Choudhury$^{2}$
\\
$^{1}$Center of Excellence in Space Sciences India, Indian Institute of Science Education and Research Kolkata, Mohanpur 741246, West Bengal, India\\
$^{2}$Inter-University Center for Astronomy and Astrophysics, Post Bag 4, Pune, Maharashtra 411007, India\\
$^{3}$UM-DAE Centre for Excellence in Basic Sciences, University of Mumbai, Vidyanagari Campus,Kalina, Santacruz (East), Mumbai, Maharashtra 400098, India
}

\date{Accepted XXX. Received YYY; in original form ZZZ}

\pubyear{2020}

\begin{document}
\label{firstpage}
\pagerange{\pageref{firstpage}--\pageref{lastpage}}
\maketitle

\begin{abstract}
We present the timing and spectral studies of RX J0209.6-7427 during its rare 2019 outburst using observations with the Soft X-ray Telescope (SXT) and Large Area X-ray Proportional Counter (LAXPC) instruments on the \textit{AstroSat} satellite. Pulsations having a periodicity of 9.29 s were detected for the first time by the \textit{NICER} mission in the 0.2-10 keV energy band and, as reported here, by \textit{AstroSat} over a broad energy band covering 0.3-80 keV. The pulsar exhibits a rapid spin-up during the outburst. Energy resolved folded pulse profiles are generated in several energy bands in 3-80 keV. To the best of our knowledge this is the first report of the timing and spectral characteristics of this Be binary pulsar in hard X-rays. There is suggestion of evolution of the pulse profile with energy. The energy spectrum of the pulsar is determined and from the best fit spectral values, the X-ray luminosity of RX J0209.6-7427 is inferred to be ${1.6}\times 10^{39}$\,ergs\,s$^{-1}$. Our timing and spectral studies suggest that this source has features of an ultraluminous X-ray pulsar in the Magellanic Bridge. Details of the results are presented and discussed in terms of the current ideas.
\end{abstract}

\begin{keywords}
RX J0209.6-7427, Be/X-ray binaries, Magellanic Bridge, type II outbursts, ULX pulsars
\end{keywords}



\section{Introduction}

Be/X-ray binary systems are a subclass of High mass X-ray binaries (HMXBs) harbouring a
compact object, usually a neutron star, and an early type massive companion star \citep{reig2011x}. The companion
Be type star has a thin Keplarian disc in the equatorial plane of the massive star formed from the matter ejected from the rapidly rotating star \citep{porter2003}. The signature of the presence of equatorial disc around the Be star manifests primarily in H $\alpha$ emission lines in the optical spectrum of the Be star \citep{slettebak1988, hanuschik1996structure, porter2003} and the infrared excess  \citep{reig2011x}. These accretion powered compact objects exhibit archetypical X-ray emission in the form of outbursts which are classified into two categories viz. type I and type II on the basis of their luminosity. Type I outbursts are less luminous (typical luminosity $< 10^{37}$ \,ergs\,s$^{-1}$), are more frequent and usually occur during the periastron passage of the compact object when it encounters the Be star disc and accretion sets in. The Type II outbursts are more luminous (typical luminosity $> 10^{37}$ \,ergs\,s$^{-1}$), less common and these outbursts are probably caused by episodes of sudden mass eruption from the Be star or due to accretion from the warped disc of the Be star which does not lie in the orbital plane of the binary system. However the exact mechanism of the eruptions is still poorly understood \citep{reig2011x, cheng2014spin, reig2018warped}. 

The X-ray source RX J0209.6-7427 was detected in the \textit{ROSAT} all sky survey and is listed in the \textit{ROSAT} PSPC catalog. \citet{kahabka2005discovery} examined the archival \textit{ROSAT} data and discovered that it had undergone two X-ray outbursts, one in March and another in October 1993 separated by a gap of $\sim$200 days. They identified it with a V=14 magnitude star and from optical spectroscopy detected a broad H $\alpha$ emission line (equivalent width of $-10.8\pm 0.2 ~\angstrom$) that suggested that the companion is a Be star and it is most likely a Be X-ray binary. From its position its location is inferred in the Magellanic Bridge connecting Small and Large Clouds \citep{kahabka2005discovery}. The outbursts in March and October 1993 lasted for about 40 days and 30 days respectively. Although there were no X-ray observations in between these two outbursts, \citet{kahabka2005discovery} detected a $\sim$39 day periodicity in the X-ray flux which they suggested may be the orbital period of the binary. However, they clarified that in view of the sparse coverage of the X-ray outburst, this needed verification. Data from the All Sky Monitor (ASM) on the \textit{Rossi X-ray Timing Explorer (RXTE)} for 1996-2011 period did not reveal occurrence of any transient at the location of RX J0209.6-7427. Similarly, \textit{INTEGRAL}, the \textit{Monitor of All Sky Image (MAXI)} and the \textit{Neil Gehrels Swift Observatory} data did not show any detectable flux from the source. Thus it is probably safe to suggest that the source remained dormant for a very long time (about 26 years) and suddenly came alive on 2019 November 20 when its X-ray activity was detected by the \textit{MAXI} mission and named as \textit{MAXI} J0206-749 as it was thought to be a new transient \citep{maxi2019atel}. This was followed up by the \textit{Neil Gehrels Swift Observatory} which located its position more precisely with the Soft X-ray Telescope (XRT) and identified \textit{MAXI} J0206-749 with RX J0209.6-7427 \citep{swift2019atel}. The optical spectrum of RX J0209.6-7427 was taken on the following day with the \textit{South African Large Telescope (SALT)} which exhibited H $\alpha$ emission having equivalent width of $-10.7\pm 0.24 ~\angstrom$ \citep{salt2019atel} which is remarkably comparable to that reported earlier by \cite{kahabka2005discovery} from the 2004 \textit{VLT/UT} observations of the likely companion Be star. Soon afterwards 9.29 s pulsations in the 0.2-12 keV band were detected from the compact object by the \textit{Neutron star Interior Composition Explorer (NICER)} mission on 2019 November 21 (\citealt{iwakiri2019atel}, also corroborated by \citealt{vasilopoulos2020}). A near-infrared source was detected at the location of RX J0209.6-7427 by the 1.4 m \textit{IRSF (InfraRed Survey Facility)} in \textit{Sutherland observatory} during observations from 2019 November 22-2019 November 25 \citep{ir2019atel}. In fact, \citet{ir2019atel} report occurrence of an infrared flare during this period. Almost simultaneous detection of H $\alpha$ emission and infrared emission during the recent X-ray outburst unequivocally suggests the presence of a disc around the companion star and corroborates that RX J0209.6-7427 is indeed a Be/X-ray binary. The Be/X-ray binaries are known to show frequent type I outbursts whenever the compact object encounters the disc around the Be star near the periastron passage. The long quiescence of this Be/X-ray binary on timescales of over two decades is very intriguing if the putative orbital period of the system is indeed $\sim$39 days or more likely $\sim$25 days as suggested by variation of the spin period discussed in
detail in Section 3.1 (see Fig. $\ref{f5}$). In addition there is a possible $\sim$47 days orbital period derived from the long source
monitoring by the \textit{Fermi} Gamma-ray Burst Monitor (GBM)
\footnote{\scriptsize{\url{https://gammaray.nsstc.nasa.gov/gbm/science/pulsars/lightcurves/rxj0209.html}}}
during the recent burst as
discussed later in the paper.\\

In this paper, we investigate the timing and spectral characteristics of RX J0209.6-7427 from soft X-rays to hard X-rays (the 0.3-80 keV band) using X-ray observations from the \textit{AstroSat} mission. Simultaneous X-ray observations from the SXT and the LAXPC instruments onboard \textit{AstroSat} space observatory were used in this study. The paper is organized as follows. After introduction, in section 2 we first describe observations from the \textit{AstroSat} mission followed by SXT and LAXPC data analysis procedures. In section 3 we describe our results related to the measurement and confirmation of the spin period of the neutron star and exploration of short-term spin period excursions in RX J0209.6-7427. We detect rapid spin-up of the
pulsar. The X-ray pulse profiles in broad energy intervals covering the 0.3-80 keV band are derived. Results from the spectral analysis covering soft
to hard X-rays in the 0.5-50 keV band, are presented. Based on the estimate of the flux, the X-ray luminosity of the source is inferred. We compare our timing and spectral results with those of other known Be binary pulsars in Small Magellanic Cloud (SMC) that underwent outbursts. Finally the results of RX J0209.6-7427 are compared with those of other known pulsating Ultraluminous X-ray sources (ULXs) in other galaxies as well as a ULX pulsar in our Galaxy. We conclude that RX J0209.6-7427
is most likely the first ULX in the SMC (Magellanic Bridge). A summary of our findings is presented in section 4.

\section{Observations and data reduction}

\begin{figure}
\centering
  \includegraphics[width=\linewidth]{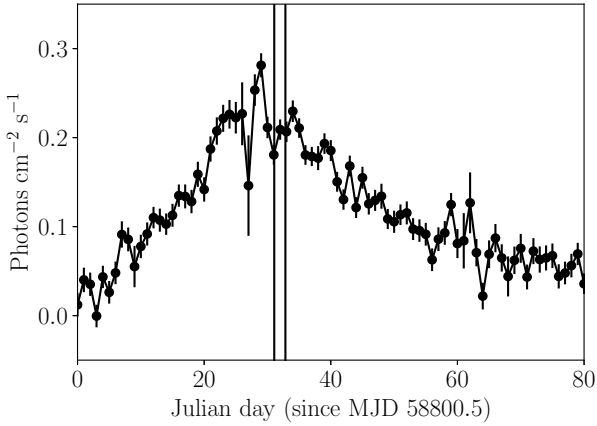} 
  \caption{\textit{MAXI} one day averaged lightcurve of RX J0209.6-7427 in the 2-20 keV energy band. The duration of overlapping \textit{AstroSat} LAXPC observations are shown by solid vertical lines.}
 \label{f1}
\end{figure}

\textit{AstroSat} Target of Opportunity (ToO) observations of RX J0209.6-7427 were performed approximately three weeks after the source woke up from its deep slumber on 2019 November 20 detected by the \textit{MAXI} mission \citep{maxi2019atel}. The X-ray observations were spread over almost 23 contiguous orbits from 2019 December 14 until 2019 December 16. In this study, we have analyzed data from the SXT and the LAXPC instruments covering
orbits 22771-22798 yielding a total exposure of about 160 ks. The log of \textit{AstroSat} observations used in our study is shown in Table ${\ref{t1}}$. The epochs of \textit{AstroSat} observations overlapping on the \textit{MAXI} light curve, are shown in Fig. \ref{f1}.

\begin{table}
\caption{\normalsize{Log of \textit{AstroSat} LAXPC observations used in this study.}} 
\label{t1}
\centering 
\begin{tabular}{c c c c c} 
\hline\hline 
\normalsize{S. no.} & \normalsize{Orbit} & \normalsize{MJD (start)} & \normalsize{Useful exposure (s)}  \\  

\hline 

\normalsize{1} & {22771} & {58831.56} & {1650}  \\
\normalsize{2} & {22772} & {58831.60} & {4870}  \\
\normalsize{3} & {22773} & {58831.67} & {5162}  \\
\normalsize{4} & {22774} & {58831.74} & {5406}  \\
\normalsize{5} & {22777} & {58831.80} & {13975}  \\
\normalsize{6} & {22778} & {58832.03} & {5302} \\
\normalsize{7} & {22780} & {58832.10} & {5291}  \\
\normalsize{8} & {22781} & {58832.17} & {5248}  \\
\normalsize{9} & {22782} & {58832.24} & {5160}  \\
\normalsize{10} & {22783} & {58832.31} & {5197}  \\
\normalsize{11} & {22784} & {58832.38} & {5438}  \\
\normalsize{12} & {22785} & {58832.47} & {5270}  \\
\normalsize{13} & {22786} & {58832.54} & {5299}  \\
\normalsize{14} & {22787} & {58832.61} & {5322}  \\
\normalsize{15} & {22788} & {58832.68} & {5339} \\
\normalsize{16} & {22789} & {58832.74} & {5414} \\
\normalsize{17} & {22791} & {58832.82} & {9712} \\
\normalsize{18} & {22792} & {58832.96} & {5362} \\
\normalsize{19} & {22793} & {58833.04} & {5311} \\
\normalsize{20} & {22795} & {58833.11} & {5219} \\
\normalsize{21} & {22796} & {58833.18} & {5186} \\
\normalsize{22} & {22797} & {58833.26} & {5116} \\
\normalsize{23} & {22798} & {58833.33} & {5052} \\

\hline 
\end{tabular}
\label{table:nonlin} 
\end{table}

\subsection{Soft X-ray Telescope}
\begin{figure}
\centering
  \includegraphics[width=0.9\linewidth]{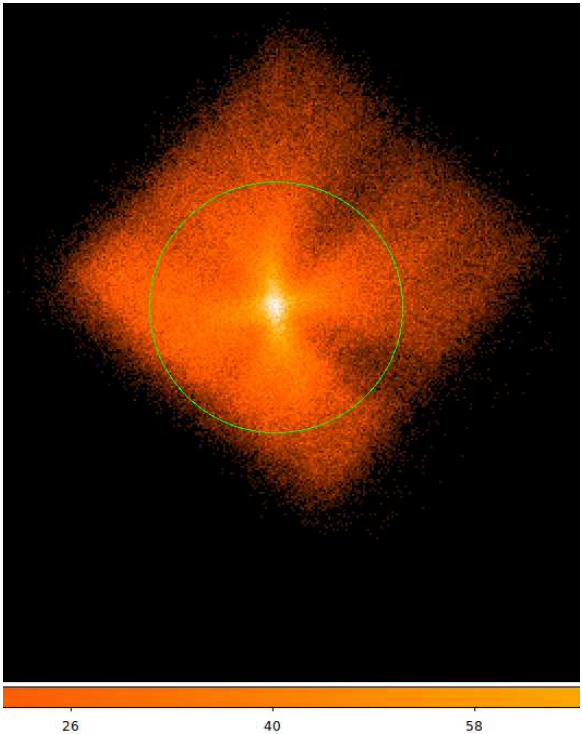} 
  \caption{SXT image of RX J0209.6-7427 in the 0.3-8 keV energy band. The circle with a 4$\arcmin$ radius, centered on the source, shows the region used for
extracting lightcurves and spectra.}
 \label{f1_new}
\end{figure}

SXT instrument is a soft X-ray telescope sensitive in the 0.3-8 keV range onboard \textit{AstroSat} satellite \citep{agrawal2006broad}. The effective area of SXT is $\sim$90 cm$^{2}$ at 1.5 keV. A detailed description of the SXT instrument can be found in \citep{singh2016orbit, singh2017soft}. The SXT observations of RX J0209.6-7427 were in the Fast window (FW) mode. The FW mode has a time resolution of $\sim$0.3 s, free from pile-up effect and especially meant to observe bright sources. The SXT level 1 data from 22771-22798 orbits were processed using SXTPIPELINE version AS1SXTLevel2-1.4b\footnote{\scriptsize{\url{http://www.tifr.res.in/~astrosat\_sxt/sxtpipeline.html}}} released on 2019 January 3, to generate level2 data for each orbit. The Level 2 SXT data of individual orbits are merged using Julia code \url{http://astrosat-ssc.iucaa.in/?q=sxtData}\footnote{\scriptsize{\url{http://www.tifr.res.in/~astrosat\_sxt/dataanalysis.html}}\label{sxt}}. The X-ray image of the source obtained with the SXT is shown in Fig. \ref{f1_new}. A circular region with 4 arcmin radius centered on the source (Fig. \ref{f1_new}), was used to extract the image, lightcurves, and spectra for the source and background respectively. 

Lightcurves and spectrum were generated using XSELECT utility in HEASoft package\footnote{\scriptsize{\url{http://heasarc.gsfc.nasa.gov/}}} (v 6.26.1). We have used SkyBkg\_comb\_EL3p5\_Cl\_Rd16p0\_v01.pha background file and sxt\_pc\_mat\_g0to12.rmf response file. Further we have used SXT arf generation tool (sxtARFModule \textsuperscript{\ref{sxt}}) to generate vignetting corrected arf  ARFTESTS1\_Rad4p0\_VigCorr.arf using the arf provided for FW mode SXT arf sxt\_fw\_excl00\_v04\_20190608.arf by the SXT instrument team.

\subsection{Large Area X-ray Proportional Counter}
   
Large Area X-ray Proportional Counter (LAXPC) instrument onboard \textit{AstroSat} mission consists of 3 identical collimated detectors (LAXPC10, LAXPC20, and LAXPC30), having 5 anode
layer geometry with 15 cm deep X-ray detection volume providing an effective area of about 4500 $cm^2$ at 5 keV, 6000 $cm^2$ at 10 keV, and about 5600 $cm^2$ at about 40 keV \citep{roy2019laxpc}.
The arrival times of X-ray photons are recorded with a time resolution of 10 $\mu$s. 
The details of the characteristics of the LAXPC instrument are available in  \citet{yadav2016}, \citet{agrawal2017large}, and \citet{roy2016performance}. The calibration details of LAXPC instrument are given in \citet{antia2017calibration}. 
We have used LaxpcSoft\footnote{\scriptsize{\url{http://astrosat-ssc.iucaa.in/?q=laxpcData}}} software (LaxpcSoft; version as of 2018 May 19) to reduce Level-1 raw data file to Level-2 data. Level-2 data contains (i) lightcurve in broad band counting mode (modeBB) and (ii) event mode data (modeEA) with information about arrival time, pulse height, and layer of origin of each detected X-ray and (iii) housekeeping data and parameter files are stored in mkf file. The standard routines available in LaxpcSoft are used to generate the light curves and the energy spectrum. The LAXPC30 detector suffered abnormal gain changes and was switched off on 2018 March 8. In the third observation (O3), the LAXPC10 detector was operating at
low gain and so we have used data only from LAXPC20 detector in our study.

\begin{figure}
\centering

\includegraphics[width=\linewidth]{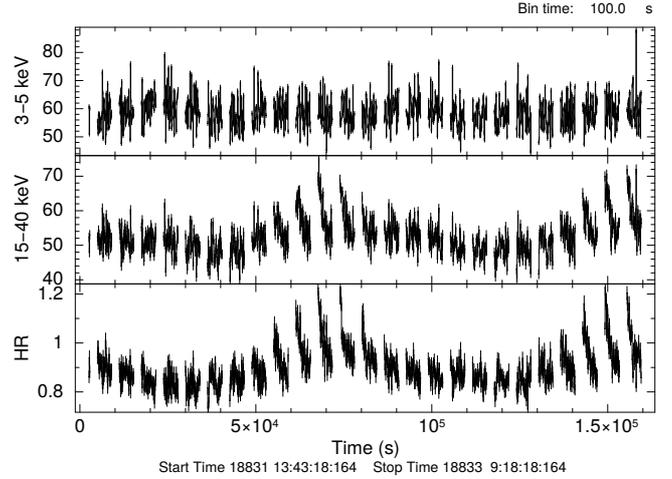} 
  \caption{Plot showing the 100 s averaged count rates from the LAXPC20 in
the soft 3-5 keV band S (top panel), in the hard 15-40 keV band H (middle panel) and the
corresponding hardness ratio (HR=H/S) shown in the bottom panel. Please note that all LAXPC lightcurves have been corrected for the background.}
 \label{f2}
\end{figure}

Fig. $\ref{f2}$ shows the lightcurves of RX J0209.6-7427 in the 3-5 keV band and the 15-40 keV band obtained by using count rates averaged over 100 s from LAXPC20 observations from orbit 22771-22798. 

Note that the LAXPC20 light curves have been corrected for the background using the routines available in the LaxpcSoft software. We correct the X-ray photons arrival times to the solar system barycenter using the \textit{AstroSat} barycentric correction utility \textquotedblleft as1bary\textquotedblright. The orbit files for barycentric correction are generated using \textit{AstroSat} orbit file generator\footnote{\scriptsize{\url{http://astrosat-ssc.iucaa.in:8080/orbitgen/}}}. \textquotedblleft as1bary \textquotedblright requires HEASoft software package (version 6.17 or higher) and so we have used the latest HEASoft software package (version 6.26) for our analysis. We show the temporal variation in the hardness ratio (HR=A/B where A and B are the count rates obtained in the energy range 15-40 keV and 3-5 keV respectively) in the third panel of Fig. $\ref{f2}$. The mean value of HR is around 0.9 with indication of
modulation around 70 ks and 150 ks. This requires further detailed investigations that we propose to
carry out in the future.\\ 

\section{Results and discussions}
\subsection{Timing studies}
\begin{figure}
        \centering
        \includegraphics[width=\linewidth]{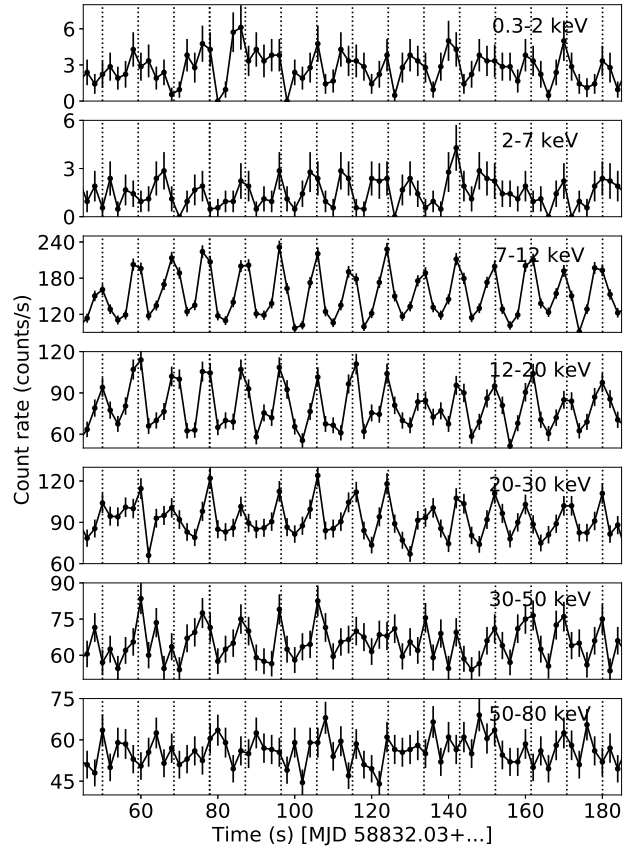}
    \caption{Energy resolved lightcurves of RX J0209.6-7427 in the 0.3-80 keV energy band using \textit{AstroSat} observations. All lightcurves have been rebinned to 2 s. Pulsations around 9 s are clearly seen in all the energy bands in the 0.3-80 keV energy range. Please note that the lightcurves shown in the 0.3-2 keV band and the 2-7 keV band are from SXT instrument while those shown in higher energy ranges (the 7-80 keV band) are from LAXPC20 observations.}
    \label{f4}
\end{figure}

The lightcurves have been generated using 2 s averaged count rates in the 0.3-2 keV band and the 2-7 keV band from SXT FW mode observations and these are shown in Fig. $\ref{f4}$. X-ray pulsations are clearly seen in the light curves. The vertical dotted lines in all the subplots indicate successive intervals of 9.28 s. Fig. $\ref{f4}$ also shows the 2 s averaged raw lightcurves in the 7-12 keV band, the 12-20 keV band, the 20-30 keV band, the 30-50 keV band, and the 50-80 keV band derived from LAXPC20 which demonstrate the presence of 9.28 s X-ray pulsations in all the energy bands indicated by the vertical dotted lines. The X-ray pulsations in hard X-rays (Energy $>$ 12 keV) are reported here for the first time in this pulsar.
We use the FTOOLS subroutine \textit{efsearch} to obtain the best estimated pulse period of RX J0209.6-7427 from the entire 160 ks LAXPC20 observations in 3-30 keV energy band. For this timing analysis we have extracted data only from the top layer of the LAXPC20 instrument to get the best signal to noise ratio. The pulse period obtained from our timing analysis is 9.28204(1) s confirming the pulse period of 9.29 s detected by the \textit{NICER} mission in soft X-rays (the 0.2-10 keV band) a few days after the source started undergoing outburst on 2019 November 20 \citep{iwakiri2019atel}. Thereafter, we investigate the temporal evolution
of the spin period of the source by estimating the pulse period for each 10 ks successive LAXPC20 observation segments. 

\begin{figure}
\centering
  \includegraphics[width=1.1\linewidth]{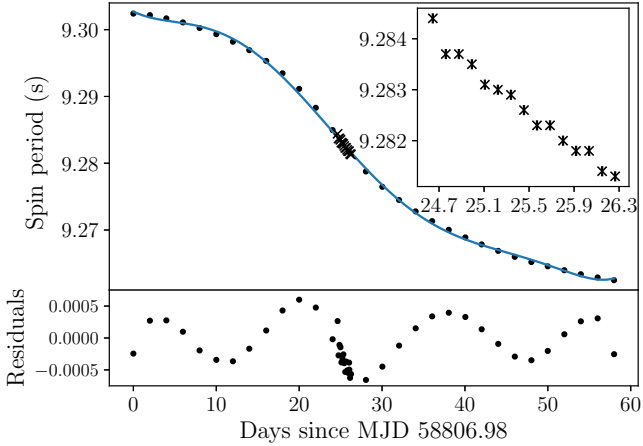} 
  \caption{Spin period evolution of RX J0209.6-7427 from \textit{FERMI/GBM} observations (filled circles) and LAXPC observations (crosses). The solid blue line shows the best polynomial fit to the spin evolution and the bottom panel shows the residuals from the poylnomial fit. The residuals in the pulse periods have a sinusoidal shape that may be due to the Doppler shift caused by the orbital motion of the neutron star. The inset shows a zoomed version of the plot showing the rapid spin-up of the pulsar inferred from spin period evolution using LAXPC observations.}
 \label{f5}
\end{figure}

The evolution of the pulsar spin period is shown in Fig. $\ref{f5}$. We have plotted the spin periods obtained from \textit{FERMI/GBM} \footnote{\scriptsize{\url{https://gammaray.msfc.nasa.gov/gbm/science/pulsars/lightcurves/rxj0209.html}}\label{fermi}} along with those obtained from our LAXPC observations. It may be noticed in the figure that the pulsar is spinning-up monotonically over the duration of \textit{AstroSat} observations spanning the period from 2019 December 14 until 2019 December 16. We fit the entire spin evolution of the pulsar using a seven degree polynomial and the residuals between the fitted model and the inferred spin periods are shown in Fig. $\ref{f5}$. It may be noted that the residuals show a sinusoidal trend having periodicity of $\sim$25 days. It is possible that the sinusoidal shape of the residuals is due to the Doppler shift of the pulsar period caused by the orbital motion of the X-ray emitting neutron star in the binary. We searched the
power density spectrum of the \textit{MAXI} one day lightcurve of this source for any periodicity due to
binary motion but did not find any dominant period. Hence origin of the sinusoidal shape residuals remains unexplained. There is a suggestion for a $\sim$47 days orbital period from the long source monitoring
by the \textit{Fermi} Gamma-ray Burst Monitor (GBM)\textsuperscript{\ref{fermi}} during the recent burst. It is possible that $\sim$47 days is a harmonic of the $\sim$25 days period.

The best fit
estimate of the spin-up rate has been derived from linear fit to the data (only the linear portion of the \textit{FERMI/GBM} spin evolution and all \textit{AstroSat} spin periods used in this fit) in Fig. $\ref{f5}$ and is $1.75 \times 10^{-8}$ \,s \,s$^{-1}$. This spin up rate is $\sim$1000 times
higher than typical value of $10^{-11}$ \,s \,s$^{-1}$ measured in typical accreting X-ray pulsar in a Be binary. This implies the accretion rate in the RX J0209.6-7427 pulsar during the outburst is about 500 to
1000 times that found in a typical pulsar. The spin-up of the pulsar during the major outburst suggests that the external accretion torque acting on the neutron star is acting in the same sense as that of the rotation of the neutron star. The rapid spin-up of the pulsar during the outburst suggests presence of an accretion disc around the neutron star \citep{ghosh1979}. It has been suggested that an accretion disc is formed during a giant outburst (type II outburst) in Be/X-ray binary systems \citep{motch1991, hayasaki2004, wilson2008} as direct accretion would be unable to explain the rapid and steady spin-up observed during type II outbursts in such sources \citep{reig2011x}.

\subsection{Broadband energy resolved pulse profiles}

\begin{figure}
        \centering
        \includegraphics[width=0.73\linewidth,angle=-90]{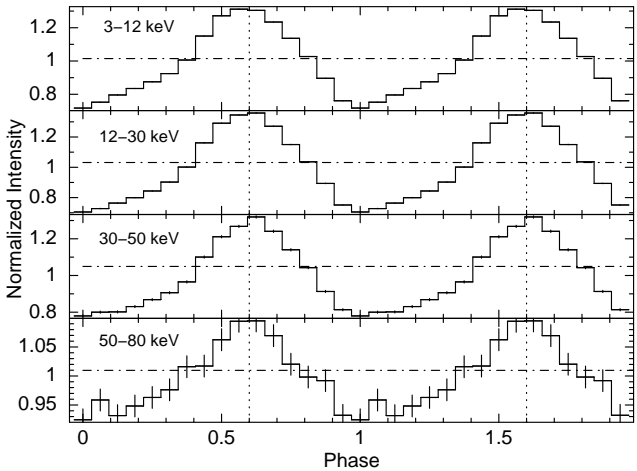}
    \caption{Folded pulse profile of RX J0209.6-7427 in the 3-12 keV band (top panel), the 12-30 keV band (second panel), the 30-50 keV band (third panel), and the 50-80 keV band (bottom panel) using LAXPC20 data. The vertical dotted lines show the location of the profile peaks in each panel while the dash-dotted horizontal lines show the half-intensity values for each folded profile.}
    \label{f6}
\end{figure}

The background subtracted folded profiles in the 3-12 keV band, the 12-30 keV band, the 30-50 keV band, and the 50-80 keV band obtained from LAXPC20 observations are shown in Fig. $\ref{f6}$.

\begin{figure}
\centering
  \includegraphics[width=\linewidth]{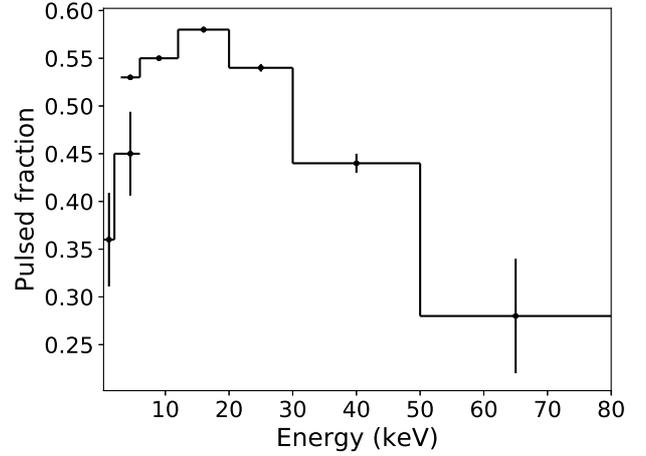} 
  \caption{Plot showing pulsed fraction in different energy ranges vs energy inferred from LAXPC20 observations.}
 \label{f7}
\end{figure}

We have extracted LAXPC20 data only from layer 1 for the 3-12 keV band and the 12-30 keV energy band while data from all the layers have been extracted for the 30-50 keV band and the 50-80 keV energy range. The energy resolved folded profiles are asymmetric especially on the leading edge and retain the profile shape with increasing energy. The variation of pulsed fraction ($\mathrm{PF}=(\mathrm{I}_{\mathrm{max}}-\mathrm{I}_{\mathrm{min}})/(\mathrm{I}_{\mathrm{max}}+\mathrm{I}_{\mathrm{min}}$)) where $\mathrm{I}_{\mathrm{max}}$ and $\mathrm{I}_{\mathrm{min}}$ are the maximum and minimum intensities in the folded profile) with energy is shown in Fig. $\ref{f7}$. 

There is indication of the pulsed fraction increasing with increasing energy in soft X-rays, reaching a maximum in the 12-20 keV band and then gradually decreasing with the increasing energy. Similar manifestation of increase in the pulsed fraction with energy has been observed in several X-ray pulsars and has been explained qualitatively using a geometrical model \citep{lutovinov2009}. However, there is suggestion of some decrease in the pulsed fraction above 30 keV which maybe due to the morphology of the accretion column of the pulsar. The asymmetric nature of folded profiles in X-ray pulsars have been suggested to occur due to the distortion of the dipole magnetic field \citep{kraus1995, sasaki2012} or higher-order multipole field components \citep{greenhill1998} or the asymmetric nature of the accretion stream near the neutron star \citep{basko1976, wang1981, miller1996}.

\subsection{Spectral studies}
\begin{figure}
\centering
  \includegraphics[width=\linewidth]{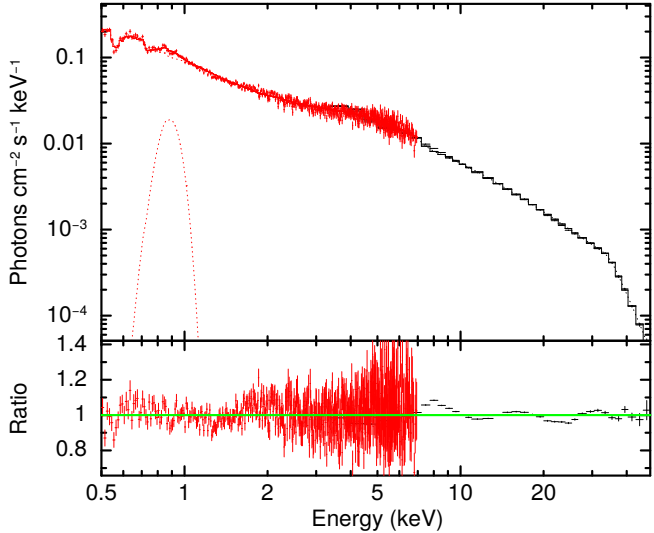} 
  \caption{Simultaneous fitted SXT and LAXPC spectrum using the power law with high energy cut off model. The best fit
mode is shown by the solid line along with the spectral data . The residuals between the data and the model are shown
in the lower panel.}
 \label{f10}
\end{figure}

We have performed a combined spectral fitting of SXT
and LAXPC20 spectra using XSPEC 12.10.1f \citep{arnaud1996astronomical} in the energy range 0.5-50 keV (Fig. $\ref{f10}$). The spectral fitting of combined SXT and LAXPC data is confined to the 0.5-50 keV band due to reliable spectral response in this energy range. A 2 percent systematics was included in the spectral analysis to take care of uncertainties in the response matrix. The broadband spectrum (the 0.5-50 keV band) is fitted using the high energy cutoff powerlaw model available in XSPEC. 

We have used a model for partial covering of partially ionised absorbing material \textit{zxipcf} \citep{reeves2008} and the $\textit{tbabs}$ model \citep{wilms2000} to take care of the broadband absorption in the spectrum during spectral fitting. An edge component has been added to fit the 4.7 keV Xe-L edge in the LAXPC spectrum with the edge position fixed at 4.7 keV. A gain correction has been applied to the SXT spectrum using the XSPEC command \textit{gain fit} where the slope was frozen at unity and the offset obtained from the fit was about 64 eV. A constant factor has been included in the model to allow for cross calibration difference between the SXT and LAXPC spectrum. We freeze the constant factor at unity for the LAXPC spectrum while allow this factor to vary for the SXT spectrum. The best
fitted constant factor for the SXT spectrum is found to be $\sim$0.9. Including a gaussian Fe-L line at 0.88 keV improved the $\chi^2$ values from $\chi^2$/d.o.f=926.34/604 to $\chi^2$/d.o.f=835.49/601. The significance of detection of this Fe-L line is 5.9 $\sigma$. The spectrum of the pulsar during the outburst is found to be hard ($\Gamma \sim 2.2$). The partial covering absorption fraction is deduced to be $\sim$0.87 with $\mathrm{n_H}$ of $16\times 10^{22
}$ \,cm$^{-2}$. It has been shown that in case of high accretion rates $\sim10^{19}$ \,g \,s$^{-1}$ in an accretion powered pulsar, an optically thick accretion envelope is formed around the ultraluminous neutron star \citep{mushtukov2017, mushtukov2019}. We surmise that the high absorption ($\sim 10^{23}$ \,cm$^{-2}$) obtained from our spectral fitting is the manifestation of the formation of dense accretion envelope around the neutron star. It is to be noted that the estimated accretion rate in RX J0209.6-7427 (see section 5) is $\sim 2.7 \times 10^{19}$ \,g \,s$^{-1}$ which satisfies the accretion rate threshold for formation of optically thick accretion envelope around an ultraluminous pulsar \citep{mushtukov2017, mushtukov2019}. The best fit spectrum gives a reduced $\chi^2$ value of 1.39 which suggests that this model can explain the continuum spectrum of the pulsar. The best fit spectral parameters of the
model are shown in Table ${\ref{t2}}$. 

\begin{table*}
\caption{\normalsize{SXT and LAXPC simultaneous spectral-fit results for RX J0209.6-7427.}} 
\label{t2}
\centering 
\begin{tabular}{lcr} 
\hline\hline 
\normalsize{
Model} & \normalsize{Parameter} & \normalsize{Value}\\  
\hline 
\normalsize{constant} & \normalsize{LAXPC spectrum}& \normalsize{1.0 (fixed)}\\
\normalsize{constant} & \normalsize{SXT spectrum} & \normalsize{$\sim 0.9$}\\
\normalsize{zxipcf} &  \normalsize{$N_{H}[10^{22}$ \,cm$^{-2}]$} & \normalsize{$16^{+0.5}_{-0.5}$}\\
 &  \normalsize{CvrFract} & \normalsize{$0.87^{+0.01}_{-0.01}$}\\
\normalsize{tbabs} & \normalsize{$N_{H}[10^{22}$ \,cm$^{-2}]$} & \normalsize{$0.22^{+0.01}_{-0.01}$}\\
\normalsize{powerlaw}    &  \normalsize{$\Gamma$} & \normalsize{$2.18^{+0.02}_{-0.02}$}\\
\normalsize{highecut}  &   \normalsize{$\mathrm{E}_{\mathrm{cut}}$[keV]} & \normalsize{$34.5^{+0.5}_{-0.5}$}\\
 & \normalsize{$\mathrm{E}_{\mathrm{fold}}$[keV]} & \normalsize{$7.9^{+0.7}_{-0.7}$}\\
\normalsize{Gaussian}  & \normalsize{$\mathrm{E(Fe-L} ~\mathrm{line})$[keV]} & \normalsize{$0.88^{+0.02}_{-0.02}$}\\
 & \normalsize{$\sigma(\mathrm{Fe-L} ~\mathrm{line})$[keV]} & \normalsize{$0.06^{+0.02}_{-0.02}$}\\
 \normalsize{Absorbed flux$^{*}$} &    \normalsize{0.5-50 ~keV} &  \normalsize{${2.52^{+0.03}_{-0.03}}$}\\
\normalsize{Absorbed luminosity$^{\#}$} &   \normalsize{0.5-50 ~keV} & \normalsize{${1.1^{+0.1}_{-0.1}}$}\\
\normalsize{Unabsorbed flux$^{*}$} &   \normalsize{0.5-50 ~keV} & \normalsize{${3.7^{+0.1}_{-0.1}}$}\\
\normalsize{Unabsorbed luminosity$^{\#}$} &   \normalsize{0.5-50 ~keV} & \normalsize{${1.6^{+0.1}_{-0.1}}$}\\
\hline 
 & \normalsize{$\chi^2$/d.o.f} & \normalsize{835.49/601} \\
 & \normalsize{${\chi^2}_{red}$} & \normalsize{1.39} \\

\hline 
\hline 
\normalsize{$^*$Flux in units of $10^{-9}$ \,ergs\,cm$^{-2}$\,s$^{-1}$}&\\
\normalsize{$^\#$Luminosity  (for a distance of 60 kpc) in units of $10^{39}$\,ergs\,s$^{-1}$}&\\

\end{tabular}
\label{table:nonlin} 
\end{table*}

From the best fit spectrum the X-ray flux of RX J0209.6-7427 in the 0.5-50 keV energy interval is deduced to be $3.7 \pm 0.1 \times 10^{-9}$ \,ergs\,cm$^{-2}$\,s$^{-1}$ which
implies X-ray luminosity to be $1.6 \pm 0.1 \times 10^{39}$\,ergs\,s$^{-1}$ for a distance of 60 kpc for the source. 
It may be noted that the absorbed flux in the soft X-rays (0.5-10 keV) inferred from our spectral analysis ($1.1 \times 10^{-9}$ \,ergs\,cm$^{-2}$\,s$^{-1}$) agrees with that derived from \textit{NICER} ($3.7 \times 10^{-10}$ \,ergs\,cm$^{-2}$\,s$^{-1}$, \citealt{iwakiri2019atel}) and that from \textit{Neil Gehrels Swift Observatory} ($3.8 \times 10^{-10}$ \,ergs\,cm$^{-2}$\,s$^{-1}$, \citealt{swift2019atel})  taking into account the increase in brightness. Indeed, the \textit{MAXI} count rates from Fig. \ref{f1} indicate that the source flux is about 2.27 times higher during the \textit{AstroSat} observations as compared to that during \textit{NICER} and \textit{Neil Gehrels Swift Observatory} observations of this source. On the other hand, the \textit{ROSAT} flux, $2.3 \times 10^{-10}$ \,ergs\,cm$^{-2}$\,s$^{-1}$, given in \citet{kahabka2005discovery}  cannot directly be compared as the energy band (0.1-2.4 keV) is different.
The inferred super Eddington luminosity is comparable to that of the 7 ULX pulsars reported so far in M82 and other galaxies. This strongly suggests that RX J0209.6-7427 is an Ultraluminous X-ray source (ULX) with an accreting neutron star with 9.29 s spin period. We do not detect any Cyclotron resonant scattering features (CRSFs) in the spectrum of this pulsar during the outburst. Usually, CRSFs are detected in the hard X-ray spectra of accretion powered pulsars during an outburst. However, some recent studies show abscence of cyclotron features in Be/X-ray binary systems. A Be/X-ray binary SMC X-3 was observed during a strong outburst \citep{tsygankov2017smc} where the inferred luminosity of the pulsar was $2.5 \times 10^{39}$\,ergs\,s$^{-1}$ but no cyclotron feature was detected. \citet{tsygankov2017smc} suggest that one of the likely reasons for the absence of cyclotron features in this ultraluminous source might be the non-dipolar nature of the magnetic field having strong multipoles. We estimate the surface magnetic field of the pulsar using the torque balance equation \citep{christodoulou2016x}
\begin{equation}
 B_*=(2 \pi^2 \xi^7)^{-1/4} \sqrt{GMI_{*}{\dot{P}_s}/{R_{*}}^6}.
\end{equation}

We use $\xi=1$, $\mathrm{M}=1.4~\mathrm{M}_{\odot}$, $\mathrm{R}_{\mathrm{*}}$=10 km, $\mathrm{I}_\mathrm{*}=2\mathrm{M}{\mathrm{R}_\mathrm{*}}^2/5$ and $\dot{\mathrm{P}}_\mathrm{s}=1.75 \times 10^{-8}$ \,s \,s$^{-1}$. The inferred surface magnetic field of the pulsar using the torque balance condition is
\begin{equation}
B=2.86 \times 10^{13} ~G.
\end{equation}

\citet{christodoulou2016x} suggest another method using minimum luminosity on the propellar line to estimate the surface magnetic field of a given pulsar using 
\begin{equation}
 B_*=8.0 \times 10^{11} \sqrt{L_X/(10^{38} ~ergs~s^{-1})/\eta} {~(P_s/1 ~s)}^{(7/6)}.
\end{equation}

 Using $\mathrm{L}_\mathrm{X}=\mathrm{L}_{\mathrm{Edd}}=1.8 \times 10^{38}$\,ergs\,s$^{-1}$ we infer the surface magnetic field to be about $2.8 \times 10^{13}$ G which
is in close agreement (within a factor of about 1.02) with that derived earlier using the torque balance method. This high field implies that any CRSF will be beyond 80 keV.

\section{Comparison of the spin evolution of RX J0209.6-7427 with those of other Be/X-ray binaries in the SMC}

In this section, we compare the observed spin-up of RX J0209.6-7427 during the 2019 giant outburst with the spin evolution in other Be/X-ray binaries in the Small Magellanic Cloud (SMC). The SMC is home of a large number of High-Mass X-ray binaries (HMXBs), most of which, excluding SMC X-1, are Be/X-ray binaries \citep{galache2008}. A catalogue of 70 Be/X-ray binaries was presented by \citet{coe2015}, while 121 relatively high-confidence HMXBs are listed in \citet{haberl2016high}. Most of the Be X-ray binaries in the SMC show spin-up during the outbursts having typical spin-up rate of about $\sim10^{-8}-10^{-11}$ \,s \,s$^{-1}$(see \citealt{galache2008} for a detailed description for each pulsar). Interestingly, the typical luminosities observed during these outbursts in all the cases except for SMC X-3 \citep{townsend2017}, are well below the Eddington limit for a neutron star accreting from a Be star. SMC X-3 showed a strong type II outburst on 2016 July 30 which was detected by the Swift observatory and the inferred luminosity of the source was about $1.2 \times 10^{39}$\,ergs\,s$^{-1}$ which exceeded the Eddington luminosity for the neutron star by a factor of about six \citep{townsend2017}. Intriguingly, SMC X-3 was spinning down before the giant outburst and then suddenly started spinning-up with spin-up rate of about $\sim 7.8 \times 10^{-10}$ \,s \,s$^{-1}$ which was almost 500 times the spin-down rate observed earlier \citep{townsend2017}. SMC X-3 may be a potential ultraluminous source when it undergoes transitions between normal accretion and ultraluminous accretion states.\\

SMC X-2 is another bright pulsar in the SMC which exhibited a bright outburst ($\mathrm{L}_\mathrm{X} \sim 10^{38}$\,ergs\,s$^{-1}$) in 2015 \citep{la2016}. However, no robust spin-up or spin-down trend could be confirmed during the outburst in this pulsar. A spin-up ($\sim5 \times 10^{-9}$ \,s \,s$^{-1}$) in this pulsar was observed during an earlier outburst detected by the \textit{RXTE} in 2000 \citep{li2016}. The Be binary J0052.1-7319 showed a 100 day long bright outburst ($\mathrm{L}_\mathrm{X} \sim8.6 \times 10^{37}$\,ergs\,s$^{-1}$) during March-October 2005 where a spin-up of about $\sim6.7 \times 10^{-9}$ \,s \,s$^{-1}$ was detected in the pulsar \citep{galache2008}. Likewise RX J0054.9-7226 exhibited a series of outbursts in mid 2002 where the measured spin-up of the pulsar was about $\sim5 \times 10^{-9}$ \,s \,s$^{-1}$ and the inferred luminosity of the source was about $\sim3 \times 10^{36}$\,ergs\,s$^{-1}$ \citep{galache2008}. One of the few known Be/X-ray binaries in the SMC undergoing spin-down during an outburst is SXP144 which showed a spin-down rate of about $\sim1.6 \times 10^{-8}$ \,s \,s$^{-1}$. The inferred luminosity of the source during the outburst was $\mathrm{L}_\mathrm{X} \sim1.1 \times 10^{36}$\,ergs\,s$^{-1}$  \citep{galache2008}.\\
\citet{ang2017} studied the evolution of luminosities and spin periods of pulsars in the SMC and suggest that pulsars having smaller spin periods P $<$ 10 s are rarely detected but usually show giant outbursts. This agrees remarkably well in the case of RX J0209.6-7427 as it has a spin period of about 9 s and shows very rare outbursts (the current outburst in this source detected after 26 years). From our comparison, we find that only SMC X-3 has shown a super Eddington outburst which is comparable to that in RX J0209.6-7427. We also note that the spin-up rates of the pulsars showing strong spin-up, are comparable to that shown by RX J0209.6-7427 despite the SMC pulsars accreting at sub-Eddington luminosities.

\section{RX J0209.6-7427: A likely new ULX pulsar in the vicinity of SMC}

A plethora of ultraluminous sources are known to exist in our Galaxy and nearby galaxies \citep{kaaret2017}, a fraction of which are believed to be accreting intermediate-mass black holes. Seven of these ULXs show X-ray pulsations demonstrating that the compact objects in them are accreting Neutron stars. Salient features of these ULX pulsars are summarised in Table ${\ref{t3}}$ along with that of RX J0209.6-7427. 

\begin{table*}
\caption{\normalsize{List of known ultraluminous X-ray pulsars.}} 
\label{t3}
\centering 
\begin{tabular}{c c c c c c c} 
\hline\hline 
\normalsize{
 Name of ULX} & \normalsize{Host Galaxy} & \normalsize{Spin period (s)} & \normalsize{Orbital period (days)} & \normalsize{Spin-up/down} & \normalsize{$L_X ~(10^{39}$\,ergs\,s$^{-1})$} & \normalsize{Reference}\\  

\hline 

 \normalsize{M82 X-2} &  \normalsize{M82} & \normalsize{1.37} & \normalsize{$\sim 2.5$} & \normalsize{Spin-up} &  \normalsize{4.9} & \normalsize{1}\\  
 \normalsize{NGC 7793 P13} &  \normalsize{NGC 7793} & \normalsize{$\sim 0.42$} & \normalsize{64} & \normalsize{Spin-up} &  \normalsize{$\sim 10$} & \normalsize{2}\\  
 \normalsize{NGC 5907 ULX} &  \normalsize{NGC 5907} & \normalsize{$\sim 1.13$} & \normalsize{5.3} & \normalsize{Spin-up} &  \normalsize{$\sim 100$} & \normalsize{3}  \\  
 \normalsize{NGC 300 ULX1} &  \normalsize{NGC 300} & \normalsize{$\sim 31.6$} & \normalsize{-} & \normalsize{Spin-up} &  \normalsize{4.7} & \normalsize{4}\\  
 \normalsize{\textit{Swift} J0243.6+6124} &  \normalsize{Milky way} & \normalsize{$\sim 9.86$} & \normalsize{$\sim 27.6$} & \normalsize{Spin-up} &  \normalsize{$\sim 2$} & \normalsize{5}\\  
\normalsize{M51 ULX-7} &  \normalsize{M51} & \normalsize{$\sim 2.8$} & \normalsize{$\sim 2$} & \normalsize{Spin-up} &  \normalsize{$\sim 10$} & \normalsize{6}\\  
 \normalsize{NGC 1313 X-2} &  \normalsize{NGC 1313} & \normalsize{$\sim 1.5$} & \normalsize{-} & \normalsize{Spin-up} &  \normalsize{$\sim 20$} & \normalsize{7}\\  
 \normalsize{RX J0209.6-7427} &  \normalsize{SMC} & \normalsize{$\sim 9$} & \normalsize{-} & \normalsize{Spin-up} &  \normalsize{$\sim 1.6$} & \normalsize{8 (this work)}\\  
\hline 
\end{tabular}
\label{table:nonlin} 
\\\normalsize{(1) \citet{bachetti2014},} (2) \citet{furst2016} and \citet{israel2017a}, (3) \citet{israel2017b}, (4) \citet{carpano2018}, (5) \citet{wilson2018}, (6) \citet{castillo2019}, (7) \citet{sathyaprakash2019}, \normalsize{and (8) this work.}
\end{table*}

Of these eight pulsating ULX sources, NGC 300 ULX1 and \textit{Swift} J0243.6+6124 are known to be in Be/X-ray binary systems. These ULX sources have typical luminosities of about  $\mathrm{L_X}\sim10^{39-40}$ \,ergs\,s$^{-1}$(ref. Table ${\ref{t3}}$) and the mechanism which powers the super-Eddington luminosities observed in these ultraluminous sources is poorly understood \citep{israel2017b}. Some of the possible mechanisms which can fuel the remarkably high luminosities in these accretion powered neutron stars are: presence of multipolar magnetic field components (\citealt{israel2017b}), magnetar like magnetic fields having B$\sim10^{14}$ G \citep{mushtukov2015} or high accretion rate with collimated tight beaming \citep{king2016ulxs}. We detect RX J0209.6-7427 during an outburst having high luminosity (Luminosity $> 10^{39}$\,ergs\,s$^{-1}$) which is rather high for a normal accretion powered pulsar (about 10 times higher). We estimate the accretion rate ($\dot{m}$) in RX J0209.6-7427 using the balance between the rate of gravitational potential energy released by the accreted matter on the surface of the neutron star and the X-ray luminosity
\begin{equation}
 \dot{m}=\frac{L_{X}R}{GM}.
\end{equation}

We use $\mathrm{L_X}={1.6}\times 10^{39}$\,ergs\,s$^{-1}$, M=1.4 $\mathrm{M}_{\odot}$ and R=10 km and obtain $\dot{m}=4.2 \times 10^{-7} ~\mathrm{M}_{\odot}$ \,yr$^{-1}$ (or $\sim 2.65 \times 10^{19}$ \,g \,s$^{-1}$) which satisfies the accretion rate threshold ($\dot{m} > 10^{-8} ~\mathrm{M}_{\odot} ~\mathrm{yr}^{-1}$; \citealt{king2016ulxs}) for RX J0209.6-7427 to be an accretion powered ULX pulsar. Some
distinguishing features of the pulsating ULXs are super-Eddington luminosities, small spin period ($\mathrm{P_{spin}} \sim$ 0.4 s to 31 s; \citealt{ray2019}), hard spectrum, and strong spin-up ($\dot{\mathrm{P}} \sim 10^{-7}-10^{-11}$ \,s \,s$^{-1}$) during an outburst \citep{trudolyubov2007chandra, bachetti2014, furst2016, israel2017a, israel2017b, carpano2018, wilson2018, castillo2019, sathyaprakash2019}. Furthermore, we find remarkable similarities in the timing and spectral features of the first ULX pulsar in the Galaxy viz. \textit{Swift} J0243.6+6124 \citep{wilson2018} and RX J0209.6-7427. It may be noted that both these ultraluminous sources are Be/X-ray binary pulsars having comparable spin periods of about 9 s. Interestingly, both these sources show giant (type II) outbursts, energy dependent profile evolution, variation in hardness ratio, dramatic spin-up showing characteristic S-shape curve, asymmetric folded profiles and lack of cyclotron features in the spectrum. It may be noted that RX J0209.6-7427 has super-Eddington luminosity, a hard power law spectrum, small spin period, and exhibits strong spin-up during a giant outburst. The remarkable similarities between some well known features of pulsating ULXs and RX J0209.6-7427 suggests that this pulsar is most likely a new ULX source in the SMC in the Magellanic Bridge.   

\section{Summary and conclusion}
We have detected broadband X-ray pulsations from RX J0209.6-7427 over a broad energy interval using the SXT and LAXPC instruments onboard the \textit{AstroSat} mission during the recent outburst of the Be/X-ray binary system in 2019. Spin-up of the pulsar is detected during the outburst from the LAXPC observations and suggests that accretion is mediated through an accretion disc around the pulsar. Broadband energy resolved pulse profiles of the pulsar have been generated. The pulse profile evolve with energy which is ubiquitous
in X-ray pulsars. The energy spectrum of the pulsar has been derived over the 0.5-50 keV band from the combined data of SXT and LAXPC. Iron-L line at 0.88 keV has been detected as shown in Fig. $\ref{f10}$. Remarkable similarities have been detected between the timing and spectral features in this luminous pulsar and those of the ULXs and suggest that this pulsar may be a ULX in the Magellanic Bridge in the vicinity of SMC.

\section*{Acknowledgements}

We are extremely thankful to the reviewer for carefully going through the manuscript  and making detailed, valuable and constructive suggestions which have greatly improved the presentation of this  paper. We are thankful to Georgios Vasilopoulos for his valuable critical comments that improved the consistency and the correctness of our result.
This publication uses the data from the \textit{AstroSat} mission of the Indian Space Research Organisation (ISRO), archived at the Indian Space Science Data Centre (ISSDC). We thank members
of LAXPC instrument team for their contribution to the
development of the LAXPC instrument. We also acknowledge
the contributions of the \textit{AstroSat} project team at ISAC. We thank the LAXPC POC at TIFR for verifying and releasing the data. LaxpcSoft software is used for analysis in this paper. This work has also used the data from the Soft X-ray Telescope (SXT) developed at TIFR, Mumbai, and the SXT POC at TIFR is thanked for verifying and releasing the data via the ISSDC data archive and providing the necessary software tools. This research has made use of software provided by the High Energy Astrophysics Science Archive Research Center (HEASARC), which is a service of the Astrophysics Science Division at NASA/GSFC. This research
has made use of the \textit{MAXI} lightcurve provided by RIKEN, JAXA, and
the \textit{MAXI} team. This research has also made use of the \textit{FERMI/GBM} pulsar spin evolution history provided by the \textit{FERMI} team. ADC is thankful to the \textit{AstroSat} Science Support Cell (ASSC) for answering queries related to the LAXPC software. ADC acknowledges support of the INSPIRE fellowship of the DST, Govt. of India. JR acknowledges ISRO for funding support and IUCAA for their facilities.




\bibliographystyle{mnras}
\bibliography{main} 

\begin{thebibliography}{}
\makeatletter
\relax
\def\mn@urlcharsother{\let\do\@makeother \do\$\do\&\do\#\do\^\do\_\do\%\do\~}
\def\mn@doi{\begingroup\mn@urlcharsother \@ifnextchar [ {\mn@doi@}
  {\mn@doi@[]}}
\def\mn@doi@[#1]#2{\def\@tempa{#1}\ifx\@tempa\@empty \href
  {http://dx.doi.org/#2} {doi:#2}\else \href {http://dx.doi.org/#2} {#1}\fi
  \endgroup}
\def\mn@eprint#1#2{\mn@eprint@#1:#2::\@nil}
\def\mn@eprint@arXiv#1{\href {http://arxiv.org/abs/#1} {{\tt arXiv:#1}}}
\def\mn@eprint@dblp#1{\href {http://dblp.uni-trier.de/rec/bibtex/#1.xml}
  {dblp:#1}}
\def\mn@eprint@#1:#2:#3:#4\@nil{\def\@tempa {#1}\def\@tempb {#2}\def\@tempc
  {#3}\ifx \@tempc \@empty \let \@tempc \@tempb \let \@tempb \@tempa \fi \ifx
  \@tempb \@empty \def\@tempb {arXiv}\fi \@ifundefined
  {mn@eprint@\@tempb}{\@tempb:\@tempc}{\expandafter \expandafter \csname
  mn@eprint@\@tempb\endcsname \expandafter{\@tempc}}}

\bibitem[\protect\citeauthoryear{Agrawal}{Agrawal}{2006}]{agrawal2006broad}
Agrawal P.,  2006, Advances in Space Research, 38, 2989

\bibitem[\protect\citeauthoryear{Agrawal et~al.,}{Agrawal
  et~al.}{2017}]{agrawal2017large}
Agrawal P.,  et~al., 2017, Journal of Astrophysics and Astronomy, 38, 30

\bibitem[\protect\citeauthoryear{Antia et~al.,}{Antia
  et~al.}{2017}]{antia2017calibration}
Antia H.,  et~al., 2017, The Astrophysical Journal Supplement Series, 231, 10

\bibitem[\protect\citeauthoryear{Arnaud}{Arnaud}{1996}]{arnaud1996astronomical}
Arnaud K.,  1996, in ASP Conf..

\bibitem[\protect\citeauthoryear{Bachetti et~al.,}{Bachetti
  et~al.}{2014}]{bachetti2014}
Bachetti M.,  et~al., 2014, Nature, 514, 202

\bibitem[\protect\citeauthoryear{Basko \& Sunyaev}{Basko \&
  Sunyaev}{1976}]{basko1976}
Basko M.,  Sunyaev R.~A.,  1976, Monthly Notices of the Royal Astronomical
  Society, 175, 395

\bibitem[\protect\citeauthoryear{Carpano, Haberl, Maitra  \&
  Vasilopoulos}{Carpano et~al.}{2018}]{carpano2018}
Carpano S.,  Haberl F.,  Maitra C.,   Vasilopoulos G.,  2018, Monthly Notices
  of the Royal Astronomical Society: Letters, 476, L45

\bibitem[\protect\citeauthoryear{Cheng, Shao  \& Li}{Cheng
  et~al.}{2014}]{cheng2014spin}
Cheng Z.-Q.,  Shao Y.,   Li X.-D.,  2014, The Astrophysical Journal, 786, 128

\bibitem[\protect\citeauthoryear{Christodoulou, Kazanas  \&
  Laycock}{Christodoulou et~al.}{2016}]{christodoulou2016x}
Christodoulou D.,  Kazanas D.,   Laycock S.,  2016, arXiv preprint
  arXiv:1606.07096

\bibitem[\protect\citeauthoryear{Coe \& Kirk}{Coe \& Kirk}{2015}]{coe2015}
Coe M.,  Kirk J.,  2015, Monthly Notices of the Royal Astronomical Society,
  452, 969

\bibitem[\protect\citeauthoryear{F{\"u}rst et~al.,}{F{\"u}rst
  et~al.}{2016}]{furst2016}
F{\"u}rst F.,  et~al., 2016, The Astrophysical Journal Letters, 831, L14

\bibitem[\protect\citeauthoryear{Galache, Corbet, Coe, Laycock, Schurch,
  Markwardt, Marshall  \& Lochner}{Galache et~al.}{2008}]{galache2008}
Galache J.,  Corbet R.,  Coe M.,  Laycock S.,  Schurch M.,  Markwardt C.,
  Marshall F.,   Lochner J.,  2008, The Astrophysical Journal Supplement
  Series, 177, 189

\bibitem[\protect\citeauthoryear{Ghosh \& Lamb}{Ghosh \&
  Lamb}{1979}]{ghosh1979}
Ghosh P.,  Lamb F.,  1979, The Astrophysical Journal, 234, 296

\bibitem[\protect\citeauthoryear{Greenhill, Galloway  \& Storey}{Greenhill
  et~al.}{1998}]{greenhill1998}
Greenhill J.,  Galloway D.,   Storey M.,  1998, Publications of the
  Astronomical Society of Australia, 15, 254

\bibitem[\protect\citeauthoryear{Haberl \& Sturm}{Haberl \&
  Sturm}{2016}]{haberl2016high}
Haberl F.,  Sturm R.,  2016, Astronomy \& Astrophysics, 586, A81

\bibitem[\protect\citeauthoryear{Hanuschik}{Hanuschik}{1996}]{hanuschik1996structure}
Hanuschik R.,  1996, Astronomy and Astrophysics, 308, 170

\bibitem[\protect\citeauthoryear{Hayasaki \& Okazaki}{Hayasaki \&
  Okazaki}{2004}]{hayasaki2004}
Hayasaki K.,  Okazaki A.~T.,  2004, Monthly Notices of the Royal Astronomical
  Society, 350, 971

\bibitem[\protect\citeauthoryear{Israel et~al.,}{Israel
  et~al.}{2017a}]{israel2017b}
Israel G.~L.,  et~al., 2017a, Science, 355, 817

\bibitem[\protect\citeauthoryear{Israel et~al.,}{Israel
  et~al.}{2017b}]{israel2017a}
Israel G.,  et~al., 2017b, Monthly Notices of the Royal Astronomical Society:
  Letters, 466, L48

\bibitem[\protect\citeauthoryear{{Iwakiri} et~al.,}{{Iwakiri}
  et~al.}{2019}]{iwakiri2019atel}
{Iwakiri} W.,  et~al., 2019, The Astronomer's Telegram, \href
  {https://ui.adsabs.harvard.edu/abs/2019ATel13309....1I} {13309, 1}

\bibitem[\protect\citeauthoryear{Kaaret, Hua  \& Roberts}{Kaaret
  et~al.}{2017}]{kaaret2017}
Kaaret P.,  Hua F.,   Roberts T.~P.,  2017, Annu. Rev. Astron. Astrophysics,
  55, 303

\bibitem[\protect\citeauthoryear{Kahabka \& Hilker}{Kahabka \&
  Hilker}{2005}]{kahabka2005discovery}
Kahabka P.,  Hilker M.,  2005, Astronomy \& Astrophysics, 435, 9

\bibitem[\protect\citeauthoryear{{Kennea} et~al.,}{{Kennea}
  et~al.}{2019}]{swift2019atel}
{Kennea} J.~A.,  et~al., 2019, The Astronomer's Telegram, \href
  {https://ui.adsabs.harvard.edu/abs/2019ATel13303....1K} {13303, 1}

\bibitem[\protect\citeauthoryear{King \& Lasota}{King \&
  Lasota}{2016}]{king2016ulxs}
King A.,  Lasota J.-P.,  2016, Monthly Notices of the Royal Astronomical
  Society: Letters, 458, L10

\bibitem[\protect\citeauthoryear{Kraus, Nollert, Ruder  \& Riffert}{Kraus
  et~al.}{1995}]{kraus1995}
Kraus U.,  Nollert H.-P.,  Ruder H.,   Riffert H.,  1995, The Astrophysical
  Journal, 450, 763

\bibitem[\protect\citeauthoryear{La~Palombara, Sidoli, Pintore, Esposito,
  Mereghetti  \& Tiengo}{La~Palombara et~al.}{2016}]{la2016}
La~Palombara N.,  Sidoli L.,  Pintore F.,  Esposito P.,  Mereghetti S.,
  Tiengo A.,  2016, Monthly Notices of the Royal Astronomical Society: Letters,
  458, L74

\bibitem[\protect\citeauthoryear{Li, Hu, Lin  \& Kong}{Li
  et~al.}{2016}]{li2016}
Li K.,  Hu C.-P.,  Lin L.,   Kong A.~K.,  2016, The Astrophysical Journal, 828,
  74

\bibitem[\protect\citeauthoryear{Lutovinov \& Tsygankov}{Lutovinov \&
  Tsygankov}{2009}]{lutovinov2009}
Lutovinov A.,  Tsygankov S.,  2009, Astronomy Letters, 35, 433

\bibitem[\protect\citeauthoryear{Miller}{Miller}{1996}]{miller1996}
Miller G.~S.,  1996, The Astrophysical Journal Letters, 468, L29

\bibitem[\protect\citeauthoryear{{Monageng}, {Buckley}, {Coe}, {Gandhi},
  {Paice}, {Charles}, {Misra}  \& {Thomas}}{{Monageng}
  et~al.}{2019}]{salt2019atel}
{Monageng} I.~M.,  {Buckley} D.,  {Coe} M.~J.,  {Gandhi} P.,  {Paice} J.~A.,
  {Charles} P.~A.,  {Misra} R.,   {Thomas} N.~T.,  2019, The Astronomer's
  Telegram, \href {https://ui.adsabs.harvard.edu/abs/2019ATel13307....1M}
  {13307, 1}

\bibitem[\protect\citeauthoryear{{Morihana}, {Mihara}, {Moritani}, {Nakajima},
  {Kawachi}, {Miyakawa}  \& {Nagayama}}{{Morihana} et~al.}{2019}]{ir2019atel}
{Morihana} K.,  {Mihara} T.,  {Moritani} Y.,  {Nakajima} M.,  {Kawachi} A.,
  {Miyakawa} K.,   {Nagayama} T.,  2019, The Astronomer's Telegram, \href
  {https://ui.adsabs.harvard.edu/abs/2019ATel13315....1M} {13315, 1}

\bibitem[\protect\citeauthoryear{Motch, Stella, Janot-Pacheco  \&
  Mouchet}{Motch et~al.}{1991}]{motch1991}
Motch C.,  Stella L.,  Janot-Pacheco E.,   Mouchet M.,  1991, The Astrophysical
  Journal, 369, 490

\bibitem[\protect\citeauthoryear{Mushtukov, Suleimanov, Tsygankov  \&
  Poutanen}{Mushtukov et~al.}{2015}]{mushtukov2015}
Mushtukov A.~A.,  Suleimanov V.~F.,  Tsygankov S.~S.,   Poutanen J.,  2015,
  Monthly Notices of the Royal Astronomical Society, 454, 2539

\bibitem[\protect\citeauthoryear{Mushtukov, Suleimanov, Tsygankov  \&
  Ingram}{Mushtukov et~al.}{2017}]{mushtukov2017}
Mushtukov A.~A.,  Suleimanov V.~F.,  Tsygankov S.~S.,   Ingram A.,  2017,
  Monthly Notices of the Royal Astronomical Society, 467, 1202

\bibitem[\protect\citeauthoryear{Mushtukov, Ingram, Middleton, Nagirner  \&
  van~der Klis}{Mushtukov et~al.}{2019}]{mushtukov2019}
Mushtukov A.~A.,  Ingram A.,  Middleton M.,  Nagirner D.~I.,   van~der Klis M.,
   2019, Monthly Notices of the Royal Astronomical Society, 484, 687

\bibitem[\protect\citeauthoryear{{Negoro} et~al.,}{{Negoro}
  et~al.}{2019}]{maxi2019atel}
{Negoro} H.,  et~al., 2019, The Astronomer's Telegram, \href
  {https://ui.adsabs.harvard.edu/abs/2019ATel13300....1N} {13300, 1}

\bibitem[\protect\citeauthoryear{Porter \& Rivinius}{Porter \&
  Rivinius}{2003}]{porter2003}
Porter J.~M.,  Rivinius T.,  2003, Publications of the Astronomical Society of
  the Pacific, 115, 1153

\bibitem[\protect\citeauthoryear{Ray et~al.,}{Ray et~al.}{2019}]{ray2019}
Ray P.~S.,  et~al., 2019, The Astrophysical Journal, 879, 130

\bibitem[\protect\citeauthoryear{Reeves, Done, Pounds, Terashima, Hayashida,
  Anabuki, Uchino  \& Turner}{Reeves et~al.}{2008}]{reeves2008}
Reeves J.,  Done C.,  Pounds K.,  Terashima Y.,  Hayashida K.,  Anabuki N.,
  Uchino M.,   Turner M.,  2008, Monthly Notices of the Royal Astronomical
  Society: Letters, 385, L108

\bibitem[\protect\citeauthoryear{Reig}{Reig}{2011}]{reig2011x}
Reig P.,  2011, Astrophysics and Space Science, 332, 1

\bibitem[\protect\citeauthoryear{Reig \& Blinov}{Reig \&
  Blinov}{2018}]{reig2018warped}
Reig P.,  Blinov D.,  2018, Astronomy \& Astrophysics, 619, A19

\bibitem[\protect\citeauthoryear{Rodr{\'\i}guez~Castillo
  et~al.,}{Rodr{\'\i}guez~Castillo et~al.}{2019}]{castillo2019}
Rodr{\'\i}guez~Castillo G.,  et~al., 2019, arXiv preprint arXiv:1906.04791

\bibitem[\protect\citeauthoryear{Roy et~al.,}{Roy
  et~al.}{2016}]{roy2016performance}
Roy J.,  et~al., 2016, Experimental Astronomy, 42, 249

\bibitem[\protect\citeauthoryear{Roy et~al.,}{Roy et~al.}{2019}]{roy2019laxpc}
Roy J.,  et~al., 2019, The Astrophysical Journal, 872, 33

\bibitem[\protect\citeauthoryear{Sasaki, M{\"u}ller, Kraus, Ferrigno  \&
  Santangelo}{Sasaki et~al.}{2012}]{sasaki2012}
Sasaki M.,  M{\"u}ller D.,  Kraus U.,  Ferrigno C.,   Santangelo A.,  2012,
  Astronomy \& Astrophysics, 540, A35

\bibitem[\protect\citeauthoryear{Sathyaprakash et~al.,}{Sathyaprakash
  et~al.}{2019}]{sathyaprakash2019}
Sathyaprakash R.,  et~al., 2019, Monthly Notices of the Royal Astronomical
  Society: Letters, 488, L35

\bibitem[\protect\citeauthoryear{Singh et~al.,}{Singh
  et~al.}{2016}]{singh2016orbit}
Singh K.~P.,  et~al., 2016, in Space Telescopes and Instrumentation 2016:
  Ultraviolet to Gamma Ray. p. 99051E

\bibitem[\protect\citeauthoryear{Singh et~al.,}{Singh
  et~al.}{2017}]{singh2017soft}
Singh K.,  et~al., 2017, Journal of Astrophysics and Astronomy, 38, 29

\bibitem[\protect\citeauthoryear{Slettebak}{Slettebak}{1988}]{slettebak1988}
Slettebak A.,  1988, Publications of the Astronomical Society of the Pacific,
  100, 770

\bibitem[\protect\citeauthoryear{Townsend, Kennea, Coe, McBride, Buckley, Evans
   \& Udalski}{Townsend et~al.}{2017}]{townsend2017}
Townsend L.,  Kennea J.~A.,  Coe M.,  McBride V.,  Buckley D.,  Evans P.,
  Udalski A.,  2017, Monthly Notices of the Royal Astronomical Society, 471,
  3878

\bibitem[\protect\citeauthoryear{Trudolyubov, Priedhorsky  \&
  C{\'o}rdova}{Trudolyubov et~al.}{2007}]{trudolyubov2007chandra}
Trudolyubov S.~P.,  Priedhorsky W.~C.,   C{\'o}rdova F.~A.,  2007, The
  Astrophysical Journal, 663, 487

\bibitem[\protect\citeauthoryear{Tsygankov, Doroshenko, Lutovinov, Mushtukov
  \& Poutanen}{Tsygankov et~al.}{2017}]{tsygankov2017smc}
Tsygankov S.,  Doroshenko V.,  Lutovinov A.,  Mushtukov A.,   Poutanen J.,
  2017, Astronomy \& Astrophysics, 605, A39

\bibitem[\protect\citeauthoryear{Vasilopoulos et~al.,}{Vasilopoulos
  et~al.}{2020}]{vasilopoulos2020}
Vasilopoulos G.,  et~al., 2020, arXiv preprint arXiv:2004.03022

\bibitem[\protect\citeauthoryear{Wang \& Welter}{Wang \&
  Welter}{1981}]{wang1981}
Wang Y.-M.,  Welter G.,  1981, Astronomy and Astrophysics, 102, 97

\bibitem[\protect\citeauthoryear{Wilms, Allen  \& McCray}{Wilms
  et~al.}{2000}]{wilms2000}
Wilms J.,  Allen A.,   McCray R.,  2000, The Astrophysical Journal, 542, 914

\bibitem[\protect\citeauthoryear{Wilson-Hodge et~al.,}{Wilson-Hodge
  et~al.}{2018}]{wilson2018}
Wilson-Hodge C.~A.,  et~al., 2018, The Astrophysical Journal, 863, 9

\bibitem[\protect\citeauthoryear{Wilson, Finger  \& Camero-Arranz}{Wilson
  et~al.}{2008}]{wilson2008}
Wilson C.~A.,  Finger M.~H.,   Camero-Arranz A.,  2008, The Astrophysical
  Journal, 678, 1263

\bibitem[\protect\citeauthoryear{Yadav et~al.,}{Yadav et~al.}{2016}]{yadav2016}
Yadav J.~S.,  et~al., 2016, in Space Telescopes and Instrumentation 2016:
  Ultraviolet to Gamma Ray. p. 99051D

\bibitem[\protect\citeauthoryear{Yang, Laycock, Christodoulou, Fingerman, Coe
  \& Drake}{Yang et~al.}{2017}]{ang2017}
Yang J.,  Laycock S.,  Christodoulou D.,  Fingerman S.,  Coe M.,   Drake J.,
  2017, The Astrophysical Journal, 839, 119

\makeatother
\end{thebibliography}








\bsp	
\label{lastpage}
\end{document}